\documentclass[aps,prl,floatfix,twocolumn,showpacs]{revtex4}
\usepackage{amssymb}
\usepackage{amsmath}
\usepackage{graphicx}

\begin{document}

\title{Coupling a quantum dot, fermionic leads and a microwave cavity on-chip}

\author{M.R. Delbecq, V. Schmitt, F.D. Parmentier, N. Roch, J.J. Viennot, G. F\`eve, B. Huard, C. Mora, A. Cottet and T. Kontos\footnote{To whom correspondence should be addressed: kontos@lpa.ens.fr}}
\affiliation{Laboratoire Pierre Aigrain, Ecole Normale Sup\'erieure, CNRS UMR 8551, Laboratoire associ\'e aux universit\'es Pierre et Marie Curie et Denis Diderot, 24, rue Lhomond, 75231 Paris Cedex 05, France}


\begin{abstract}
We demonstrate a hybrid architecture consisting of a quantum dot circuit coupled to a single mode of the electromagnetic field.  We use single wall carbon nanotube based circuits inserted in superconducting microwave cavities. By probing the nanotube-dot using a dispersive read-out in the Coulomb blockade and the Kondo regime, we determine an electron-photon coupling strength which should enable circuit QED experiments with more complex quantum dot circuits.
\end{abstract}

\date{\today}
\maketitle

An atom coupled to a harmonic oscillator is one of the most
illuminating paradigms for quantum measurements and amplification\cite{Clerk:10}. Recently,
the joint development of artificial two-level systems and high finesse microwave resonators in superconducting
circuits has brought the realization of this model on-chip\cite{Wallraff:04,Chiorescu:04}. This "circuit Quantum Electro-Dynamics" architecture allows, at least in principle, to combine circuits with an arbitrary complexity. In this context, quantum dots can also be used as artificial atoms\cite{Yoshie:04,Reithmaier:04}. Importantly, these systems often exhibit many-body features if coupled strongly to Fermi seas, as epitomized by the Kondo effect.
Combining such quantum dots with microwave cavities would therefore enable the study of a new type of coupled fermionic-photonic systems.

 Cavity quantum electrodynamics\cite{Raimond:01} and its electronic counterpart circuit quantum electrodynamics\cite{Clerk:10} address the interaction of
 light and matter in their most simple form i.e. down to a single photon and a single atom (real or artificial). In the field
 of strongly correlated electronic systems, the Anderson model follows the same purified spirit\cite{Georges:96}. It describes a single electronic
 level with onsite Coulomb repulsion coupled to a Fermi sea. In spite of its apparent simplicity, this model allows to capture
 non-trivial many body features of electronic transport in nanoscale circuits. It contains a wide spectrum of physical phenomena
 ranging from resonant tunnelling and Coulomb blockade to the Kondo effect. Thanks to progress in nanofabrication techniques, the
 Anderson model has been emulated in quantum dots made out of two dimensional electron gas\cite{GGordon:98}, C60 molecules\cite{Roch:08} or carbone nanotubes\cite{Cobden:02}.
 Here, we mix the two above situations.  We couple a quantum dot in the Coulomb blockade or in the Kondo regime to a single mode of the electromagnetic field
 and take a step further towards circuit QED experiments with quantum dots.

 \begin{figure}[!pth]
\centering\includegraphics[height=0.85\linewidth,angle=0]{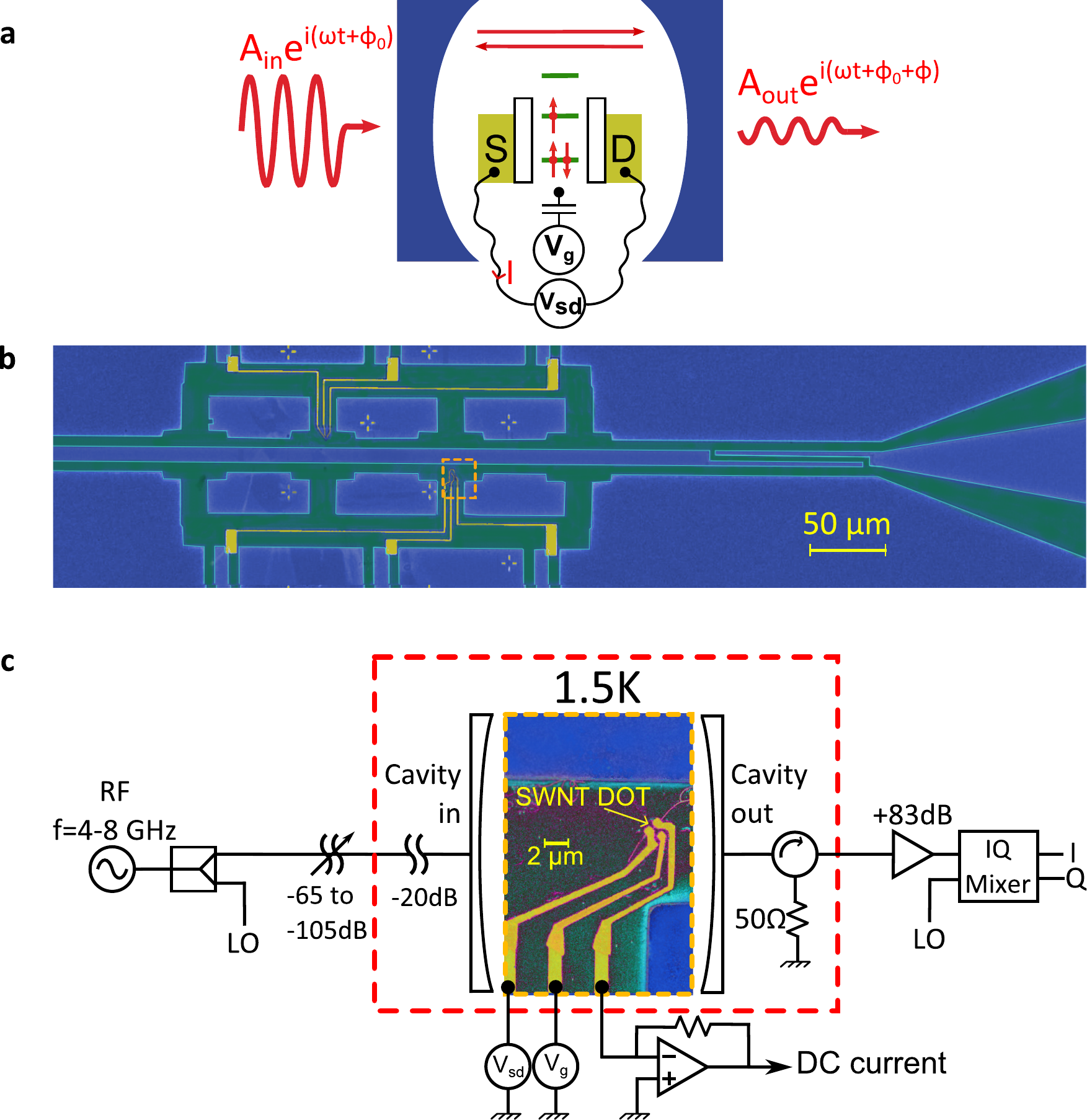}
\caption{\textbf{a.} Schematics of the quantum dot embedded in the microwave cavity. The transmitted microwave field has different amplitude and phase from the input field as a result of its interaction with the quantum dot inside the cavity. The quantum dot is connected to "wires" and capacitively coupled to a gate electrode in the conventional 3-terminal transport geometry. \textbf{b.} Scanning electron microscope (SEM) picture in false colors of the coplanar waveguide resonator. Both the typical coupling capacitance geometry of one port of the resonator and the 3-terminals geometry are visible. \textbf{c.} False colours SEM picture of a SWNT dot inside an on-chip cavity embedded in a schematics of the measurement setup.
}%
\label{device}
\end{figure}

Low frequency charge transport in quantum dots in the Coulomb blockade or Kondo regime has been studied with exquisite details\cite{Cobden:02,Grobis:08}. However, their dynamic
aspects have remained to a great extent unexplored so far. Previous studies have tackled the problem in terms of photo-assisted electron tunnelling\cite{Elzerman:00,Kogan:04}. Here, we focus on the dispersive effect of the quantum dot on the microwave field.
In order to enhance the electron-photon interaction which would be otherwise too small to be detected, we place our quantum dot circuit inside
an on-chip microwave cavity as depicted in figure 1a. One important aspect of our approach is the implementation of "wires" which go
inside the cavity (see figure 1). A source (S) and a drain (D) electrode are used to drive a DC current through the  quantum dot. A gate
electrode (G) is used to control in situ the position of the energy levels on the dot. At the same time, a microwave continuous signal
in the 4-8GHz range is sent to one port of the cavity and amplified through the other port. Both quadratures of the transmitted signal
are measured. The temperature of the experiment is 1.5K. As shown in figure 1b and 1c, we use single wall carbon nanotubes (SWNTs) embedded in superconducting microwave on-chip cavities in order to implement the model situation of figure 1a. SWNTs are ideally suited
to implement the kind of experiments we discuss here. They can be contacted with normal\cite{Cobden:02}, superconducting\cite{Buitelaar:02,Cleuziou:06,Herrmann:10} or ferromagnetic\cite{Cottet:06,Feuillet:10}
materials to form various kinds of  hybrid systems. Here, we investigate the most simple case i.e. a single quantum dot connected to
two normal metal leads and capacitively coupled to a side gate electrode, as shown in figure 1c. However, our scheme can readily be generalized to more complex circuits like double quantum dots.

The phase of the microwave signal transmitted through the cavity is particularly sensitive to the presence of the quantum dot circuit.
Figure 2a displays the color scale plot of the low frequency differential conductance of one particular device as a function of the
source-drain voltage $V_{sd}$ and the gate voltage $V_g$. We observe the characteristic "Coulomb diamonds" with resonant lines in the $V_{sd}-V_g$
plane as well as the characteristic "Kondo ridge" at zero bias from $V_g= -2.5V$ to $V_g= -2.0V$,  signalling the emergence of the Kondo
effect. As shown in figure 2c in black line, the conductance for $V_g= -2.32V$ peaks up to $0.75 \times 2e^2/h$, which is close to its maximum
possible value $2e^2/h$. The corresponding variations of the phase of the microwave signal in the vicinity of the cavity resonance, at
4.976 GHz, are displayed as a function of $V_{sd}$ and $V_g$ in the color scale plot of figure 2b. Essentially all the spectroscopic features
observed in the conductance are visible in the phase spectroscopy. In particular, a similar peak at zero bias as in the DC conductance
is observed as shown in red line in figure 2c. It corresponds to a variation of about $2.10^{-3} rad$ which is not proportional to the DC conductance in general as shown in figure 2c.

\begin{figure}[!pth]
\centering\includegraphics[height=0.65\linewidth,angle=0]{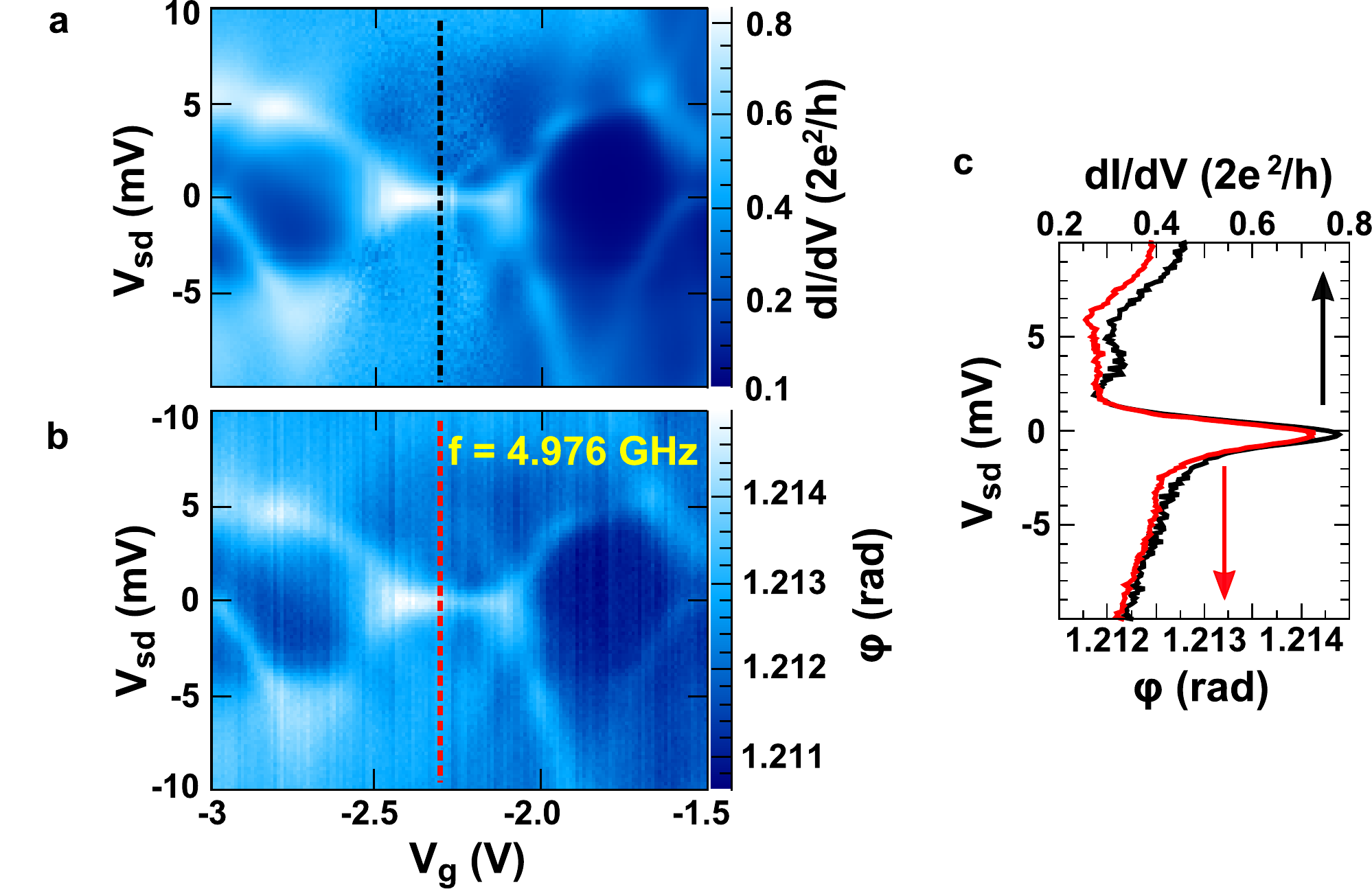}
\caption{\textbf{a.} Color scale plot of the differential conductance in units of $2e^2/h$ measured along three charge states exhibiting the conventional transport spectroscopy. A Kondo ridge is visible at zero bias around $V_g = -2.3V$. \textbf{b.} Color scale plot of the phase of the microwave signal  at f = 4.976 GHz, measured simultaneously with the conductance of figure 2a. \textbf{c.} Differential conductance and phase of the transmitted microwave signal at f = 4.976 GHz as a function of source drain bias $V_{sd}$ for $V_g=-2.32V$.
}%
\label{greyscale}%
\end{figure}

The observation of the Kondo resonance in the phase of the microwave signal shows that the fermionic and photonic systems are coupled. Our Kondo dot-cavity system has to be described by an extension of the Anderson model, known as the Anderson-Holstein model which has been devised to treat quantum impurities coupled to phonons. Our "photonic" Anderson-Holstein hamiltonian reads :  $H=H_{dot}+H_{cav}+(\lambda_K \hat{N}_K+\lambda_{K'}\hat{N}_{K'})(\hat{a}+\hat{a}^{\dagger})$ with $\lambda_{K(K')}$  and $\hat{N}_{K(K')}$  respectively, the electron-photon coupling constant and the number of electrons for each K(K') orbital of the nanotube-dot (which arise from the band structure of nanotubes), $\hat{a}$  being the photon field operator. The coupling constants $\lambda_{K(K')}$ arise from the capacitive coupling of the nanotube energy levels to the central conductor of the cavity. The terms $H_{dot}$ and $H_{cav}$  are the standard Anderson hamiltonian of a single energy level coupled to fermionic reservoirs and the standard hamiltonian of a single photon mode coupled to a photonic bath. As shown in figure 3a, the capacitive coupling between the cavity and the dot induces oscillations of the electronic level. There is also an indirect coupling through oscillations of the bias between source and drain, as indicated by the dashed lined edges of the Fermi seas in figure 3a. The resonator allows to probe both the dispersive and the dissipative response of the dot. Therefore, both the frequency and the width in energy of the bosonic mode are affected by the mutual interaction between the electronic and photonic systems. Since our cavity contains a large number of photons (about $10000$ at $-60dBm$ of input power), it is justified to use classical electrodynamics to describe the coupled systems. The circuit element corresponding to the quantum dot has a complex admittance $Y_{dot}(\omega )$, following the spirit of the scattering theory of AC transport in mesoscopic circuits\cite{Buttiker:93,Buttiker:96}.  To leading order with respect to the energy scales of the dot, one gets $Y_{dot}(\omega)\approx  \alpha/R_{dot} + j C_{dot}\omega  $.  The dissipative part is proportional to the differential conductance $1/R_{dot}$ of the dot and stems from the residual asymmetric AC coupling of the leads S and D to the cavity. The reactive part $C_{dot}$ corresponds to a capacitance. As shown in figure 3a, we model the resonator as a discrete LC-circuit with a damping resistor R (in red) coupled via coupling capacitors (in green) to external leads. The corresponding frequency broadening and frequency shift read  $\delta f_D \approx \alpha  /(2 C_{res} R_{dot})$ and  $\delta f_R \approx -C_{dot} f_0 /(2 C_{res})$ respectively, where $f_0$ is the resonance frequency and $C_{res}$ is the capacitance of the resonator.
Figure 3b shows how to directly measure  $\delta f_D$ and  $\delta f_R$. The top panel displays the expected variations of the phase close to a single resonance when a finite  $\delta f_D$ or  $\delta f_R$ are included (in red and blue respectively). The reference curve (for  $\delta f_D = \delta f_R =0$) is in black dashed lines. The lower panel shows that, subtracting the reference curve, a finite  $\delta f_D$ affects the odd part of the phase contrast curve (in red) whereas  $\delta f_R$ affects its even resonant part (in blue). From these curves,  $\delta f_R$ and  $\delta f_D$ can be directly measured from the area of the blue curve and the area of half of the red curve, respectively. The corresponding experimental curves are shown in figure 3c for $V_g=-2.44V$ (on the Kondo ridge), taking the point $V_g=-1.85V$ and Vsd=0mV as a reference. We observe a resonance at $4.976 GHz$ with a quality factor of about $160$ for the even part in blue. The oscillations of the odd part in red correspond to residual imperfections of our amplification line. We measure directly  $\delta f_R$ and  $\delta f_D$ by integrating the whole blue curve and half of the red curve (the positive part)

\begin{figure}[!pth]
\centering\includegraphics[height=0.45\linewidth,angle=0]{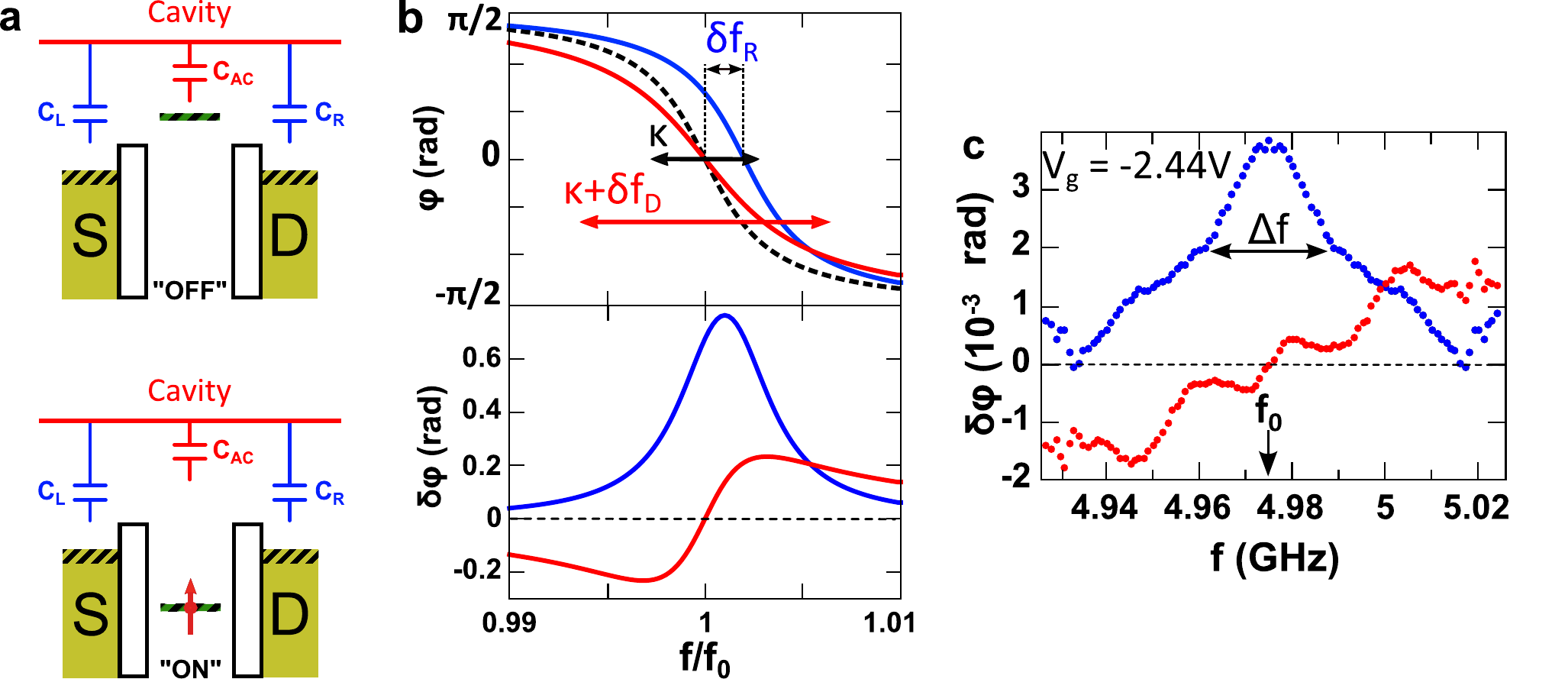}
\caption{\textbf{a.} Capacitive coupling of the quantum dot to the cavity. Both the fermionic leads and the quantum dot are coupled to the resonator, resulting in an AC modulation of both $V_{sd}$ and $V_g$ (shadings). \textbf{b.} Upper panel, the signature of dispersion and dissipation to frequency dependence of the microwave signal phase for a standard resonance. Reference resonance (black dotted), shifted by  $\delta f_R$ (blue) as a result of dispersion and broadened by  $\delta f_D$ (red) as a result of dissipation. Lower panel :  the even part (in blue) and odd part (in red) as a function of frequency. The area under the blue curve is proportional to  $\delta f_R$ and the area of half the red one is proportional to  $\delta f_D$. \textbf{c.} Even and odd parts of the phase contrast $\delta \phi$ as a function of frequency on the coulomb peak at $V_g=-2.44V$ on the spectroscopy of figure 2. The even part exhibits a resonance centered on $f = 4.976 GHz$. The odd part shows residual modulation due to imperfection in the measurement lines. }
\label{diagram}
\end{figure}

We now focus on $C_{dot}$. This quantity is a direct measurement of the charge susceptibility of the electronic system. For a single particle resonance with width $\Gamma$ , the scattering theory\cite{Buttiker:93,Buttiker:96} predicts $C_{dot} = 2e^2/\pi \Gamma $  at resonance, which amounts to re-expressing the spectral density of the single energy level coupled to the fermionic leads in terms of a quantum capacitance. If electron correlations are present, the situation changes. In the Coulomb blockade regime as well as in the Kondo regime, on expects a reduction of the capacitance on a peak with respect to that of a single particle resonance with the same width \cite{Lee:11,Delbecq:12}.

%When the dot level is far from the Fermi energies of the reservoirs, a charge can be transferred from source S to drain D only via a virtual process in which the quantum dot is doubly occupied. The %spin of the electron is reverted without changing the orbital. This process is singular at low temperature and leads to the Kondo resonance. If more than one orbital is involved %\cite{Jarillo:05,Delattre:09}, the orbital degree of freedom can be different in the final state than in the initial state. Whereas both processes yield a finite current in the Kondo regime, only the %latter yields a finite $C_{dot}$ \cite{Lee:11} since the orbital is different between the initial and final state. Therefore, $C_{dot}$  should be smaller in general for a Kondo resonance than for a %single particle resonance with the same parameters.

%the correlation induced energy renormalization of the level also change the way the occupation of the dot evolves as the gate voltage is swept.

\begin{figure}[!pth]
\centering\includegraphics[height=0.65\linewidth,angle=0]{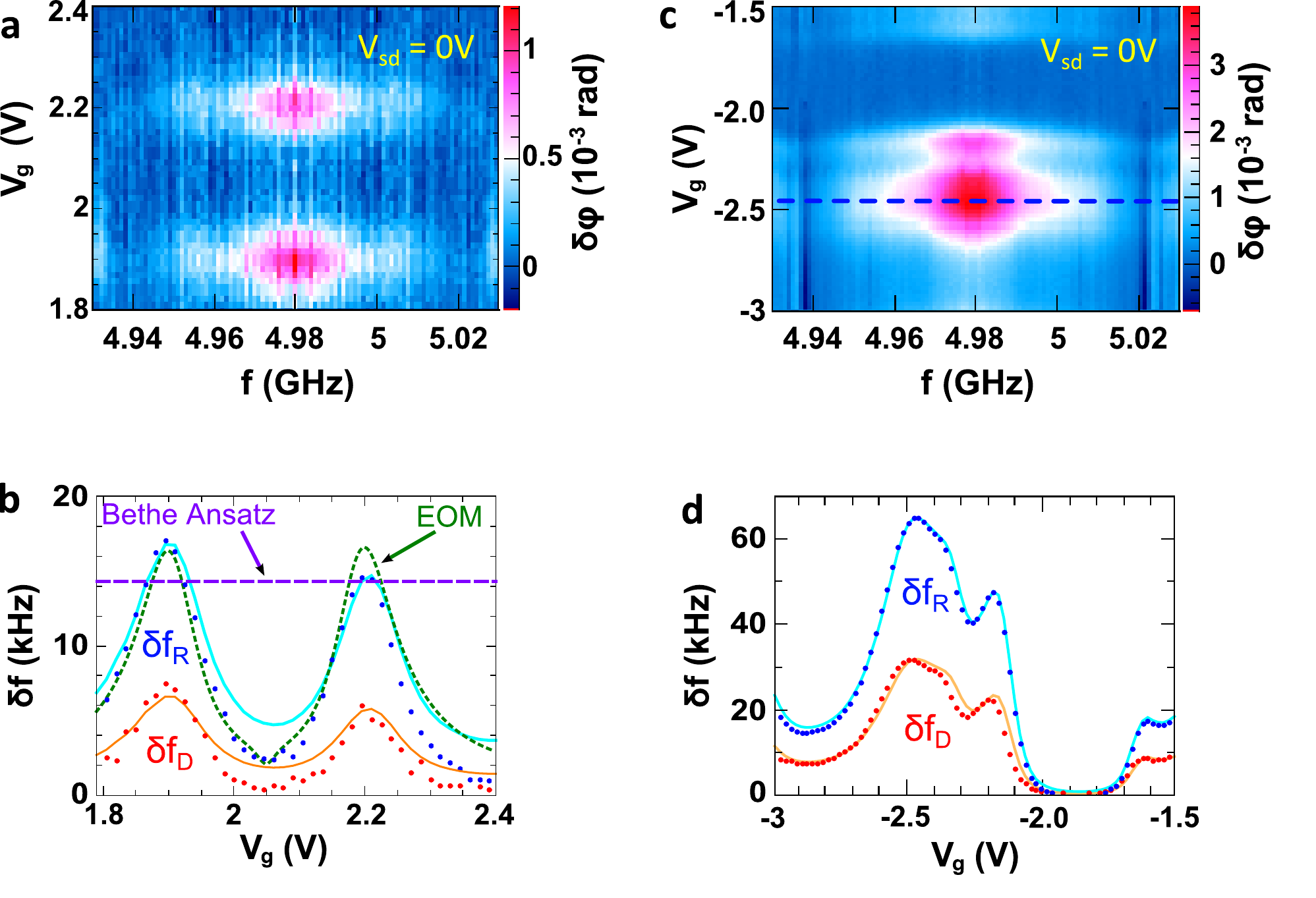}
\caption{\textbf{a.} Color scale plot of the even part of the phase contrast $\delta \phi$ of two Coulomb peaks as a function of the gate voltage $V_g$ and the frequency of the microwave signal in the vicinity of the cavity resonance.  $\delta \phi$  is taken with respect to a reference phase in the empty orbital at $V_g = 2.4V$. \textbf{b.} Gate dependence of the reactive (blue dots) and the dissipative (red dots) parts of the dot response extracted respectively from the area under the even part (figure 4a) and the area under half of the corresponding odd part. Formulae of the main text for  $\delta f_R$ (blue line) and  $\delta f_D$ (orange line)give $C_0 = 18 aF$,  $\alpha = 0.003$. Comparison with EOM theory (dashed dark green) and Bethe ansatz (dashed purple). \textbf{c.} Color scale plot of the even part of the phase contrast $\delta \phi$ of the Kondo spectroscopy shown in figure 2 as a function of the gate voltage $V_g$ and the frequency f. The line cut corresponds to the curves of figure 3c. \textbf{d.} Gate dependence of the reactive (blue dots) and the dissipative (red dots) parts of the dot response extracted respectively from the area under the even part (figure 4c) and the area under half of the corresponding odd part. Formulae of the main text for  $\delta f_R$ (blue line) and  $\delta f_D$ (orange line) give $C_0 = 22aF$,  $\alpha = 0.004$
}%
\label{coupling}%
\end{figure}

The measured even part of the phase contrast as a function of frequency and gate voltage are presented in figure 4a and c in color scale. We investigate both the Coulomb blockade (left panels) and the Kondo regime (right panels) for the same device by tuning it in different gate regions. The point at $V_g=2.4V$ ($V_g=-1.85V$) and Vsd=0mV is our phase reference for the Coulomb blockade and the Kondo regime respectively. The Coulomb blockade peaks (transport spectroscopy not shown) are visible as two elongated pink spots in the $f-V_g$ plane centered at 4.976GHz which span over 50 MHz. The measured  $\delta f_R$ and  $\delta f_D$ are shown in figure 4b in blue and red dots respectively. They modulate like Coulomb blockade peaks up to $15 kHz$ and $5kHz$ respectively. The dispersive shift  $\delta f_R$ can be directly translated into a capacitance from $f_0=4.976GHz$ and $C_{res}= 0.7 pF$, which are known from our setup. A comparison with the scaled conductance is shown in blue line using the expression $C_0 f_0 /(2 C_{res})\times dI/dV \times h/2e^2$ for $\delta f_R$ with $C_0=18aF$. Whereas the Coulomb peaks are well taken into account, this empirical formula fails to account for the Coulomb valley. The electron-photon coupling strength can be directly evaluated from these measurements. Indeed, the expected capacitance change for the quantum dot can be calculated using an Equation of Motion technique (EOM) for the Green's functions. It can also be evaluated using the Bethe Ansatz on the Coulomb peaks at $T=0$. Therefore, the measured capacitance change $\Delta C_{dot}$ of the dot is directly related to the calculated $C_{th}$ one by $\Delta C_{dot} = \alpha^{2}_{AC} C_{th}$  \cite{Delbecq:12}. The couplings $\lambda_{K(K')}$ in our on-chip Anderson-Holstein Hamiltonian can be calculated from $\lambda_{K(K')}= e \alpha_{AC} V_{rms} $. In the above expression, $V_{rms}$  corresponds to the rms voltage of a single photon in the cavity mode\cite{Wallraff:04} and $e$ to the elementary charge. As shown on figure 4b, the EOM theory, in green dashed lines, accounts well for our measurements and agrees well on the peaks with the Bethe ansatz result \cite{Delbecq:12}. From this, we extract $\alpha_{AC}\approx 0.3$,  which leads to $\lambda_{K(K')}\approx 140MHz$. The Kondo ridge of figure 2 is visible as two merged elongated pink spots. The corresponding measured  $\delta f_R$ and  $\delta f_D$ are shown in figure 4d. They both modulate up to $30 kHz$ and $60 kHz$ respectively as the gate voltage sweeps the energy levels of the dot. In particular, we extract $C_{dot}$ of 16aF for the Kondo ridge at $V_g= -2.32V$.  This allows us to provide another estimate for $\lambda_{K(K')}$  from  $\lambda_{K(K')} \approx e V_{rms} \sqrt{C_{dot}/C_{Kondo}}$.  We use  $C_{Kondo}= 4e^2/\pi T_K \approx 200aF$ as the upper bound of the capacitance expected for the Kondo ridge, $T_K$ being the full width at half maximum of the Kondo peak as measured from figure 2C. Consistently with the previous estimate, we get $\lambda_{K(K')}\approx 140MHz$. As expected\cite{Schoelkopf:98},  $\delta f_D$ is well accounted for with $ \alpha /(2R_{dot}C_{res})$, with $ \alpha =0.004$, using the measured $dI/dV=1/R_{dot}$ (see orange line in figure 4d, we present a similar curve in figure 4b for $\alpha=0.003$ ).  Interestingly, the empirical formula shown in blue line for $C_0=22aF$ is in better agreement with the measured $\delta f_R$ in the Kondo regime than in the Coulomb blockade regime. Even though this might arise from non-universal features of the Anderson Hamiltonian, the observation of a finite $C_{dot}$ is consistent with the participation of the K and K' orbitals which naturally lead to the high Kondo temperature observed here. Like for singly occupied closed double quantum dots \cite{Cottet:11}, a finite capacitance resembling the conductance is expected if $\lambda_K \neq \lambda_{K'}$  due to the finite orbital susceptibility of the dot in the Kondo regime\cite{Nishikawa:10}.

In conclusion, our method can be generalized to many other types of hybrid quantum dot circuits\cite{Trif:08,Cottet:10, Jin:11}. The measured coupling is similar to the ones demonstrated recently in superconducting circuits and can readily be used to probe the quantum regime for the microwave cavities. Generally, our findings pave the way to circuit quantum electrodynamics with complex open quantum circuits. They could be used for example to "simulate" on-chip other aspects of the Anderson-Holstein hamiltonian like polaronic effects.

\begin{acknowledgments}
We are indebted with M. Devoret,  A. Clerk, R. Lopez, R. Aguado, P. Simon, A. Levy Yeyati and G. Zarand for discussions. We gratefully acknowledge the Meso group of LPS Orsay for illuminating discussions on resonant detection techniques and P.  Bertet and L. Dumoulin for helping us to fabricate the superconducting  resonators. The devices have been made within the consortium Salle Blanche Paris Centre. This work is supported by the ANR contracts DOCFLUC, HYFONT and  SPINLOC.
\end{acknowledgments}

\end{document}